\documentclass[reprint,onecolumn,amsmath,amssymb,notitlepage]{revtex4-1}
\pdfoutput=1

\expandafter\let\csname equation*\endcsname\relax 
\expandafter\let\csname endequation*\endcsname\relax 
\usepackage{graphicx}
\usepackage{hyperref}

\makeatletter
\hypersetup{
	    pdfcreator={Cesar Henrique Comin},
		colorlinks=true,       		
    	linkcolor=black,          	
    	citecolor=black,        		
    	filecolor=black,      		
		urlcolor=black,
		bookmarksdepth=4
}
\makeatother

\begin{document}

\title{Random walks in directed modular networks}

\author{Cesar H. Comin$^1$}
\email{email: chcomin@gmail.com}
\author{Matheus P. Viana$^1$}
\author{Lucas Antiqueira$^2$}
\author{Luciano da F. Costa$^1$}
\affiliation{$^1$Instituto de F\'{\i}sica de S\~{a}o Carlos, Universidade de S\~{a}o Paulo, S\~{a}o Carlos, S\~ao Paulo, Brazil\\
$^2$Instituto de Ci\^{e}ncias Matem\'{a}ticas e de Computa\c{c}\~{a}o, Universidade de S\~{a}o Paulo, S\~{a}o Carlos, S\~ao Paulo, Brazil}

\begin{abstract}

Because diffusion typically involves symmetric interactions, scant attention has been focused on studying asymmetric cases. However, important networked systems underlain by diffusion (e.g. cortical networks and WWW) are inherently directed. In the case of undirected diffusion, it can be shown that the steady-state probability of the random walk dynamics is fully correlated with the degree, which no longer holds for directed networks. We investigate the relationship between such probability and the inward node degree, which we call \emph{efficiency}, in modular networks. Our findings show that the efficiency of a given community depends mostly on the balance between its ingoing and outgoing connections. In addition, we derive analytical expressions to show that the internal degree of the nodes do not play a crucial role in their efficiency, when considering the Erd\H{o}s-R\'enyi and Barab\'asi-Albert models. The results are illustrated with respect to the macaque cortical network, providing subsidies for improving transportation and communication systems.

\end{abstract}

\maketitle

\section{Introduction}

Diffusion is one of the most fundamental dynamics in physics. In addition to being ubiquitous, diffusion also underlies important non-linear systems such as Turing reaction-diffusion, Schr\"odinger, Fokker-Planck, and Navier-Stokes equations. Because these systems typically present symmetric interactions, few works have addressed the asymmetric counterpart. However, important networked systems in biology (e.g. cortical and metabolic networks), transportation (e.g. city traffic), and communications (e.g. WWW and email networks) are typically directed and underlain by diffusion~\cite{boccaletti:2006}. A particularly important property of such systems is how the asymptotic probability of the random walk dynamics (which we call activation) can be predicted from intrinsic topological features. For instance, it is known that the activation is fully correlated with the node degree for undirected networks~\cite{vespignani:2008}. However, this is no longer ensured in the case of diffusion in directed networks \cite{antiqueira:2011}. The present work reports an investigation on how the balance of directed edges in modular networks defines how activation increases with the inward node degree inside each community, a concept that we call \emph{efficiency}. Though general, our findings are illustrated with respect to a cortical network \cite{kaiser:2006b}, as recent studies have shown that diffusion can be used to model disease progression in the cortex \cite{raj:2012}.

\subsection{The undirected case}

There are many studies concerning diffusion over undirected networks \cite{lovasz:1993,noh:2004,samukhin:2008,bray:1988,gfeller:2007,jespersen:2000,burda:2009}, but the most recognized property of this dynamics is that its steady-state probability is completely defined by the number of neighbors (or degree) $k_i$ of a given node $i$ \cite{noh:2004}. To illustrate this, we consider a random walk taking place on the network. If the walker is at node $j$ at time $t-1$, we can write the probability of finding this agent at node $i$ at time $t$ as

\begin{equation}
    {\cal P}_i(t) = \sum_{j=1}^N\frac{A_{ij}}{k_j}{\cal P}_j(t-1),
    \label{eq:pundirected}
\end{equation}
where $k_j$ denotes the degree of node $j$, $N$ is the total number of nodes, and $A_{ij}$ is the element $(i,j)$ of the adjacency matrix $\mathbf{A}$. After a long period of time, the system is guaranteed to reach equilibrium~\footnote{The limit exists if the network contains an odd loop \cite{noh:2004}.} and we have ${\cal P}_i(t) = {\cal P}_i(t-1) = {\cal P}_i^{\infty}$. The time required to reach equilibrium may depend on degree-degree correlations \cite{gallos_2008}, but here we consider such correlations as negligible. In order to find ${\cal P}_i^{\infty}$ we can use the detailed balance condition ${\cal P}_i^{\infty}k_j={\cal P}_j^{\infty}k_i$ \cite{noh:2004}, to obtain \cite{vespignani:2008}

\begin{equation}
    {\cal P}_{i}^{\infty}=\frac{k_{i}}{N\bar{k}}\label{eq:p_undirected}.
\end{equation}
Here $\bar{k}$ means the average of $k$. For brevity and to evoke a broader physical meaning, for the remainder of this paper we call the long time probability ${\cal P}_{i}^{\infty}$ the \emph{activity}, ${\cal A}_i$, of node $i$. Muchnik et al. \cite{muchnik_2013} have studied a similar concept of activity, defined by the number of posts, messages or page edits in the Wikipedia (http://www.wikipedia.org) and a collaborative news-sharing web-site (http://www.news2.ru). They showed that the average node degree conditioned on its activity follows a smooth curve, which can be fitted with great precision by a geometric distribution. Nevertheless, here we are interested on the random walk activity, which certainly is mainly influenced by the node degrees.

\subsection{The directed case}

In the case of directed networks, the relation between the activity and the degree of a given node is not linear and remains not completely understood. In this work we focus on the study of some interesting properties that emerge when considering the direction of the edges. Fortunato et al.~\cite{fortunato:2009} studied the problem of searching in directed networks and found an expression for the PageRank dynamics, which is basically a random walk process with additional random jumps occurring with probability $q$. In the limit $q\rightarrow 0$, the PageRank reduces to our definition for the activity of a node in a strongly connected network, that is,

\begin{equation}
    {\cal A}_i=\frac{1}{N\bar{k}^{\textnormal{in}}}k_{i}^{\textnormal{in}}
    \label{eq:p_directed}
\end{equation}
where $k^{\textnormal{in}}$ is the number of connections that point to a node, called inward degree.

\begin{figure}[!tb]
    \includegraphics[width=0.9\linewidth]{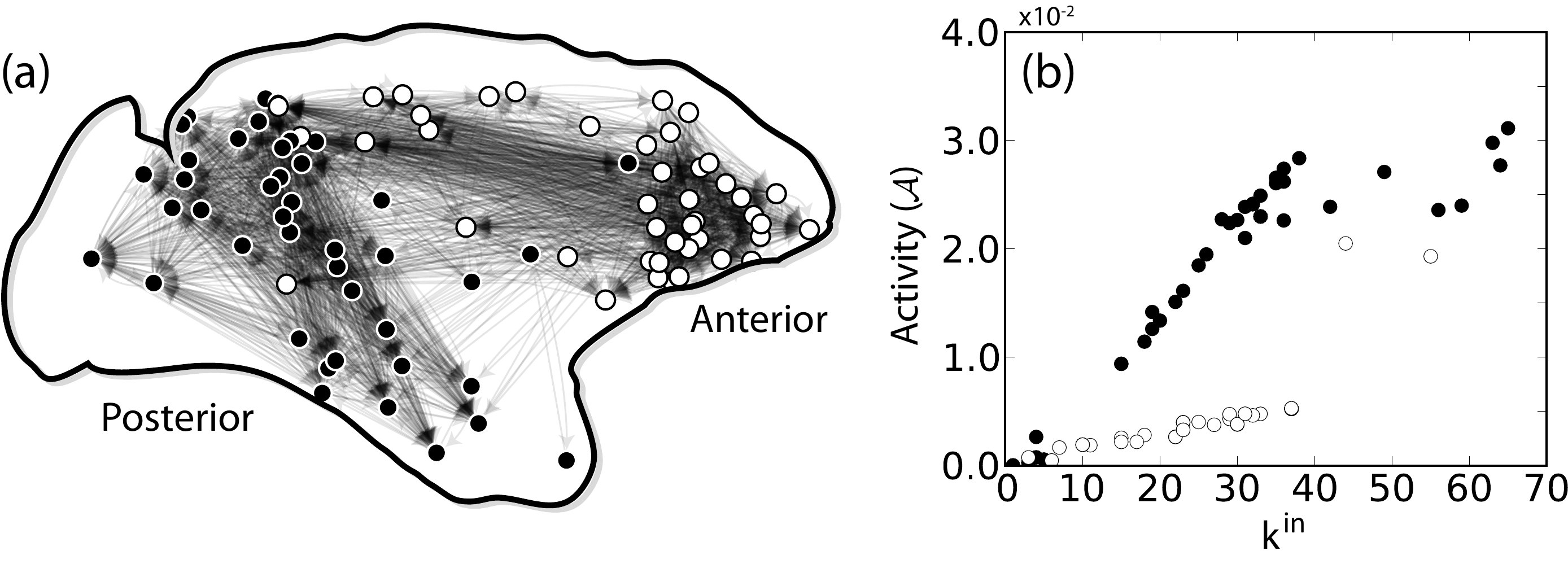}
    \caption{(a) Largest strongly connected component of the Macaque cortical network partitioned in two communities. This component has 85 nodes and average degree $\bar{k}=27.7$. (b) Relationship between the activity and inward degree of each node. Nodes in (b) are colored according to the respective community in (a).}
    \label{f:cortical}
\end{figure}

An example of a directed network where this relation can be studied is the cortical network of the Macaque of the family \emph{Cercopithecidae} \cite{kaiser:2006b} shown in Figure \ref{f:cortical}(a). In this network, each node corresponds to a cortical region and the directed edges correspond to corticocortical tracts connecting these regions. This network has been largely studied in order to understand how stimulus in specific regions of the cortex can induce activation in other regions. Concerning the topological organization of this cortical network, we observe the presence of two communities~\footnote{In order to detect the communities of the directed cortical network, we used the algorithm proposed by Leicht et al. \cite{newman:2008}}, which are mostly related to the spatial position (anterior and posterior) of the nodes, as seen in Figure \ref{f:cortical}(a). The anterior community corresponds to the regions responsible for performing high level tasks such as sensory integration and planning, while the posterior community is related to the primary sensory processing, including the cortical visual area in the occipital, temporal and parietal lobes~\cite{nsome:2005}. We also found an asymmetry of connections between these two communities, with the posterior module receiving more connections from the anterior module than the other way around. In Figure \ref{f:cortical}(b) we show ${\cal A}_i$~\footnote{Obtained through the diagonalization of the transition matrix of the random walk.} in terms of $k_i^{\textnormal{in}}$ for every node of this network, which contradicts the linear dependence between these variables suggested by Equation \ref{eq:p_directed}. The non-linear behavior shown in Figure \ref{f:cortical}(b) was first observed in Ref. \cite{antiqueira:2011}, and is here verified to be related to the modular organization of the cortical network, identified by the black and white nodes, which correspond to the posterior and anterior communities in Figure \ref{f:cortical}(a), respectively.

\section{Methods}

How does the topology of the cortical network originate the observed non-linearity? In order to address this question, we first need to consider the probability of the walker being at one of the  communities. Suppose we have a network of $N$ nodes and $n$ communities. For simplicity's sake, we also suppose that all communities have the same size $N/n$ (the case for different sizes is straightforward). The evolution of the probability $P_m$ of finding the walker at community $m$, is given by

\begin{equation}
	\partial_t P_{m}=-P_{m}\sum\limits_{h\neq m}p_{mh}+\sum\limits_{h\neq m}P_{h}p_{hm}\label{eq:master}
\end{equation}
where $p_{mh}$ is the probability of the walker going from community $m$ to $h$. In the equilibrium, $\partial_t P_m=0$ and the probability of the walker being at each community is given by the eigenvector  $\mathbf{\pi}$ associated to the unitary eigenvalue of the transition matrix ${\bf T}$, whose elements are equal to $p_{mh}$. In order to find an expression for $p_{mh}$, we group the nodes of the network in ``equivalence classes", i.e., we define the vector class ${\bf X}_{m}=(x_{m1},x_{m2},\dots,x_{mn})$ as the set of all nodes of community $m$ that makes $x_{m1}$ connections with community 1, $x_{m2}$ with community 2 and so on. Then, the elements of matrix ${\bf T}$ can be written as

\begin{equation}
p_{mh}=\sum\limits_{{\bf X}_m}P(m\rightarrow h\mid {\bf X}_m)P({\bf X}_m)\label{eq:pij}\end{equation}
where $P(m\rightarrow h\mid {\bf X}_{m})$ is the conditional probability that the walker, being in some node with exactly $x_{m1},...,x_{mn}$ connections, will move from community $m$ to $h$. This probability is clearly

\begin{equation}
    P(m\rightarrow h\mid {\bf X}_m)=
    \frac{x_{mh}}
    {x_m^{\textnormal{out}}},
    \label{eq:transition}
\end{equation}
where $x_m^{\textnormal{out}}=\sum_{l=1}^n x_{ml}$ corresponds to the outward degree of nodes inside class ${\bf X}_m$. Regarding Poissonian networks (e.g. Erd\H{o}s-R\'enyi random graph), Equation \ref{eq:pij} can be exactly solved, but since the node degrees of these networks are known to be well represented by the average value of the associated degree distribution, it is reasonable to make the homogeneous assumption and consider every node as having the same number of connections pointing to each community. Therefore, every node inside community $m$ makes $x_{mh}=\bar{x}_{mh}$ connections with community $h$ (or equivalently, every node of community $m$ is in the same class ${\bf X}_m$). Under this assumption, the probability $p_{mh}$ is related to the mixing parameter \cite{fortunato:2009b}, and can be written as 

\begin{equation}
    p_{mh}=\frac{\bar{x}_{mh}}{\bar{x}_{m}^{\textnormal{out}}}.
    \label{eq:pij_degree}
\end{equation}
Finally, we can write the activity of a node $i$ located inside community $m$ as

\begin{equation}
    {\cal A}_{i\mid m}=\frac{\pi_{m}}{\frac{N}{n}\bar{x}_m^{\textnormal{in}}}k_{i}^{\textnormal{in}}=\xi_m k_{i}^{\textnormal{in}},
    \label{eq:activity_n}
\end{equation}
where, $\bar{x}_m^{\textnormal{in}} = \sum_{l=1}^n\bar{x}_{lm}$ is the average inward degree of community $m$. Equation \ref{eq:activity_n} corresponds to the asymptotic probability of the walker being inside community $m$, $\pi_m$, times the activity of a node in a directed network, given by Equation \ref{eq:p_directed} evaluated over the nodes inside community $m$. The coefficient $\xi_m$ is here called \emph{efficiency} of the community $m$, since it represents the capacity of the community $m$ to generate activity through the inward connections of its nodes. If we were to change the relative efficiency of two given communities, we have only two options, either increase the internal average degree of one community, or change the balance between them, i.e., change the relation $\bar{x}_{mh}-\bar{x}_{hm}$. We will show that only the second option can actually change the communities efficiency.

In order to test the accuracy of Equation \ref{eq:activity_n}, we use a simple model for modular directed networks where it is possible to control the efficiency of each community. First, we define the $n\times n$ matrix $\mathbf{\kappa}$, which specify the desired internal (diagonal entries) and external (off diagonal entries) average degrees. The elements of this matrix are given by $\bar{x}_{mh}$. We also specify the reciprocity vector ${\bf r}$, which controls, for each community, the probability that an edge between two given nodes is mutual \cite{Newman:2010}. Next, we create $n$ undirected Erd\H{o}s-R\'enyi networks with $E$ edges and assign a random direction for every edge $(i,j)$. In addition, a reciprocal edge $(j,i)$ is added to the $m$-th network with  probability $r_{m}$. The final average internal degree of each network is then $\bar{x}_{mm}=nE(1+r_{m})/N$. Finally, we connect the created networks among themselves, but now considering the off-diagonal elements of $\mathbf{\kappa}$.

\section{Results}
\subsection{Efficiency for ER networks}

\begin{figure}[!tb]
    \includegraphics[width=\linewidth]{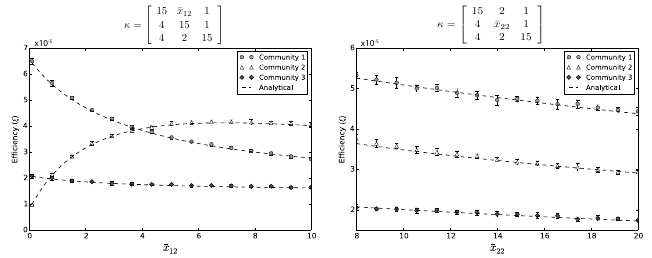}
    \caption{(a) Illustration of how the unbalance between communities implies in large efficiency differences. (b) Invariance of the relative efficiency when the balance is kept and the internal density of community 2 is changed. The dashed lines correspond to the predictions obtained from Equation \ref{eq:activity_n}. The respective $\mathbf{\kappa}$ matrix is shown above each graph. Each point represents the mean over 100 realizations of the model and the error bars show the standard deviation.}
    \label{f:3_com}
\end{figure}

Considering the model described in the previous section, in Figure \ref{f:3_com}(a) we show the communities efficiency for the case $n=3$ and $N=1500$ as a function of the average number of connections between communities 1 and 2, $\bar{x}_{12}$. For each point in this figure, we created 100 networks using the proposed model. Then, through linear regression, we estimated the angular coefficient of the relation between activity and inward degree of the nodes for each network. The dashed lines represent the value of $\xi_m$ expected by Equation \ref{eq:activity_n} for each community, found by diagonalization of the $3\times 3$ transition matrix ${\bf T}$ and using the \emph{a priori} parameters defining each community. It is clear that the homogeneous assumption works well and can correctly predict the communities efficiency.

The most striking feature of Equation \ref{eq:activity_n} is that each community can have different efficiencies. It is possible to show that if the communities are balanced, which is the case for all undirected networks, the internal degree cannot change the relative efficiency between communities. In order to show that, let us consider the case where all the communities are balanced and, consequently, $\bar{x}_{mh}$=$\bar{x}_{hm}$ for all $h\neq m$, implying $\bar{x}_{m}^{\textnormal{out}}=\bar{x}_{m}^{\textnormal{in}}$. Therefore, $\pi_m=\frac{1}{n}\bar{x}_{m}^{\textnormal{in}}/\bar{x}^{\textnormal{in}}$ is an eigenvector of the transition matrix ${\bf T}$ (which is verified by substitution), and

\begin{equation}
    {\cal A}_{i\mid m}=\frac{1}{N\bar{x}^{\textnormal{in}}}k_{i}^{\textnormal{in}}
    \label{eq:balanced}
\end{equation}
which does not depend on the community and recovers the previous result given in Equation \ref{eq:p_directed}. This result means that denser communities have greater activity, but the relation between activity and inward degree remains the same for every community. If we try to increase the activity of community $m$ by adding edges inside it, both its average inward degree $ \bar{x}_{m}^{\textnormal{in}}$ and probability $\pi_m$ will increase in the same proportion, making the efficiency nearly the same according to Equation~\ref{eq:activity_n}. The only change comes from the increase in the average inward degree $ \bar{x}^{\textnormal{in}}$ of the entire network, which in general cases is hardly affected by the change of only one community. We observed that such behavior is also verified when the communities are unbalanced. This case is shown in Figure \ref{f:3_com}(b), where we vary the internal average degree of community 2. We also verified that the reciprocity of the communities does not have any effect in their efficiency.

To better understand the effect of unbalancing the connections between communities, we start with $n$ balanced communities, and invert a fraction $1-\epsilon$ of the outgoing connections between the community $f$ and $m$. This is repeated for every community $m$, so as to greatly increase the activity of $f$. We need to show that under reasonable conditions the efficiency of $f$ after the rewiring process ($\xi^{,}_f$) is higher than the rest of the network, i.e. $\xi^{,}_f/\xi^{,}_m>1$ for a given $m$. In order to do so we first observe that the new transition probabilities, $p'_{mh}$, are given by

\begin{equation}
    p'_{mh}=\begin{cases}
    \frac{\bar{x}_{mh}+\bar{x}_{mh}(1-\epsilon)\delta_{f,h}}{\bar{x}_{m}^{\textnormal{out}}+\bar{x}_{mf}(1-\epsilon)} & m\neq f \\
    \frac{\bar{x}_{fh}\epsilon}{\epsilon\sum\limits_{l\neq f}\bar{x}_{fl}+\bar{x}_{ff}} & m=f,h\neq f \\
    \frac{\bar{x}_{ff}}{\epsilon\sum\limits_{l\neq f}\bar{x}_{fl}+\bar{x}_{ff}} & m=h=f,
    \end{cases}\label{eq:new_pij}
\end{equation}
where $\delta_{f,h}=1$ if $f=h$ and 0 otherwise. Therefore, in the limit $\epsilon \rightarrow 0$, the ratio $\pi^{,}_f/\pi^{,}_m$ for $m\neq f$, between the elements of the eigenvector $\mathbf{\pi}^{,}$ of the new transition matrix, must diverge. On the other hand, assuming that $\bar{x}_{m}^{,\textnormal{in}}$ denotes the average inward degree of nodes inside community $m$ after the inversion process, the inward degree ratio, given by

\begin{equation}
    \frac{\bar{x}_{m}^{,\textnormal{in}}}{\bar{x}_{f}^{,\textnormal{in}}}=\frac{\bar{x}_{m}^{\textnormal{in}}-(1-\epsilon)\bar{x}_{mf}}{\bar{x}_{f}^{\textnormal{in}}+(1-\epsilon)\underset{l\neq f}{\sum}\bar{x}_{fl}}\approx \frac{\bar{x}_{m}^{\textnormal{in}}-\bar{x}_{mf}}{2\bar{x}_{f}^{\textnormal{in}}-\bar{x}_{ff}}\label{eq:new_degree}
\end{equation}
will remain bounded, making the efficiency ratio between communities $f$ and $m$ being necessarily higher than one.

\subsection{Efficiency for Barab\'asi-Albert communities}

\begin{figure}[!t]
    \includegraphics[width=0.5\linewidth]{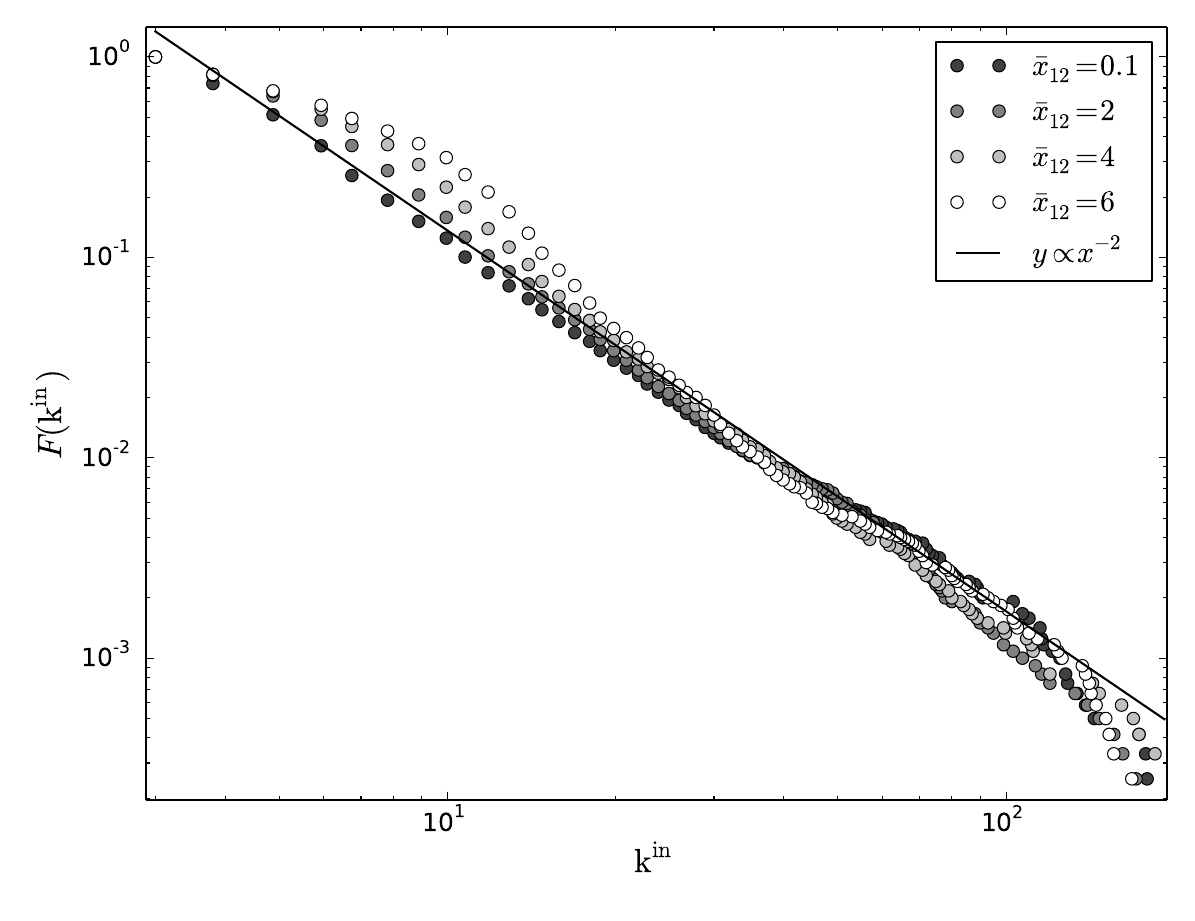}
    \caption{Cumulative degree distributions of the networks generated from the model. Refer to the text and the figure legend for details about the parameters used. We expect a relationship of the form $y\approx x^{-2}$, which is shown as a solid line in the figure.}
    \label{f:BA_degree}
\end{figure}

In the case of communities with power-law topology \cite{fortunato:2009b}, it is harder to find a closed expression for the activity, but there is nothing indicating that the same effect cannot be observed. There is also no standard model to generate a modular network with power-law topology in which the balance of connections between the communities can be systematically controlled. Nevertheless, we can define a simple model where $n$ Barab\'asi-Albert (BA) networks \cite{barabasi:1999} are created and randomly connected in the same manner as in our previously presented model. In Figure \ref{f:BA_degree} we show the cumulative degree distribution for the networks generated by this model with the following $\kappa$ matrix 

\begin{equation}
\kappa = \begin{bmatrix}
          6.0 & \bar{x}_{12} & 0.01 \\
          0.5 & 6.0 & 0.01 \\
          0.5 & 0.5 & 6.0
         \end{bmatrix}.\label{eq:kappa}
\end{equation}
The parameter $\bar{x}_{12}$ was set to the values indicated in the figure. There is a slight deviation for lower-degree nodes. This deviation increases with $\bar{x}_{12}$, but should not represent a problem since we are interested in cases where $\bar{x}_{12}< \bar{x}_{mm}$.

Using Equation \ref{eq:pij} we can write the transition probabilities of the random walk in this model as

\begin{equation}
    p_{mh}=a\prod_{\substack{
   k=1..n \\
   k\neq m}}
   \left(\sum\limits_{x_{mk}}B^{(k)}\right)\sum\limits_{x_{mm}}\frac{1}{x_{mm}^3}\frac{x_{mh}}{x_m^{\mathrm{out}}},\label{eq:prob_BA}
\end{equation}
where $B^{(k)}$ is a binomial distribution with mean $\bar{x}_{mk}$ and success probability $\frac{n\bar{x}_{mk}}{N}$. The constant $a$ is a normalization factor of the power-law distribution of $x_{mm}$. The summation terms represent the probability of finding a node belonging to the equivalence class $\mathbf{X}_m$, while the term $\frac{x_{mh}}{x_m^{\mathrm{out}}}$ is the transition probability associated to the equivalence class. For example, if the network contains three communities, the transition probability from community 1 to community 2 can be written as

\begin{equation}
    p_{12}=a\sum\limits_{x_{12}}B^{(2)}\sum\limits_{x_{13}}B^{(3)}\sum\limits_{x_{11}}\frac{1}{x_{11}^3}\frac{x_{12}}{x_{11}+x_{12}+x_{13}}\label{eq:prob_BA_three}.
\end{equation}

\begin{figure}[!t]
    \includegraphics[width=\linewidth]{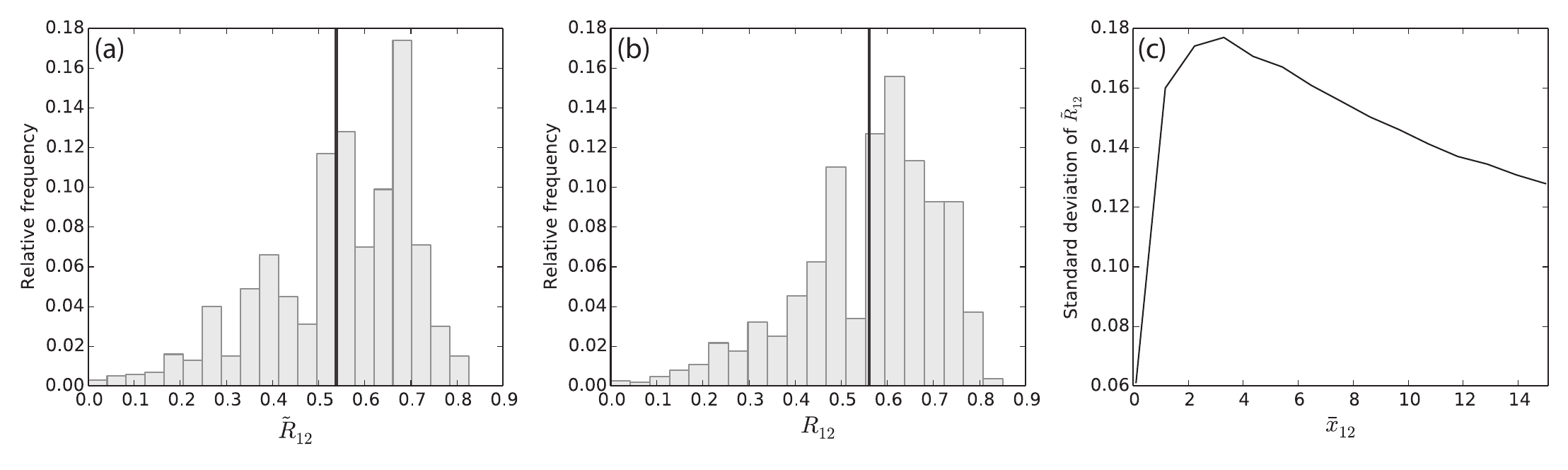}
    \caption{Distribution of transition probabilities for nodes in community 1 of a network possessing three communities. (a) Measured transition probabilities for a network generated by our model. (b) Expected transition probabilities indicated by Equation \ref{eq:prob_BA}. Vertical solid lines show the average of the values. (c) Standard deviation of the measured transition probabilities for distinct mixing between communities.}
    \label{f:deviation}
\end{figure}

Equation \ref{eq:prob_BA} can represent with good accuracy the individual node transition probabilities occurring in the network. In order to show this, we define the quantity

\begin{equation}
    R_{mh}(i)=\frac{x_{mh}(i)}{x_m^{\mathrm{out}}(i)}\label{eq:R}
\end{equation}
for each node $i$ of community $m$. The average of $R_{mh}$ over community $m$ of a network realization should be as close as possible to the value $p_{mh}$ defined in Equation \ref{eq:prob_BA}. Also, Equation \ref{eq:prob_BA} can only provide a good prediction of the community efficiency if the values $R_{mh}$ are well represented by the average $p_{mh}$. In Figure \ref{f:deviation} we show the result of the analysis regarding transition values for networks generated from the model. Only transitions from community 1 to community 2 were considered, but the other transitions display similar results. In Figure \ref{f:deviation}(a) we show the distribution of measured transition values, which we call $\tilde{R}_{12}$, from a model network created with parameters

\begin{equation}
\kappa = \begin{bmatrix}
          6.0 & 6. & 0.01 \\
          0.5 & 6.0 & 0.01 \\
          0.5 & 0.5 & 6.0
         \end{bmatrix}.
\end{equation}
We also indicate the average value of $\tilde{R}_{12}$ by a solid line. In Figure \ref{f:deviation}(b) we show the predicted distribution of transition values obtained using the term $   a\prod_{\substack{
   k=1..n \\
   k\neq m}}
   \left(\sum\limits_{x_{mk}}B^{(k)}\right)\sum\limits_{x_{mm}}\frac{1}{x_{mm}^3}$ from Equation \ref{eq:prob_BA}. The distributions, and consequently their average values, are very similar. We tested different parameters for the $\kappa$ matrix and the experimental and analytical values are always in good agreement. As noted above, another necessary condition for the model to work is that the distribution of $R_{12}$ is well represented by its average value. One way to verify this is by looking at the standard deviation of $R_{12}$, which we measured from networks generated by the model. The result is shown in Figure \ref{f:deviation}(c) for distinct values of $\bar{x}_{12}$. The other parameters are the same as for the network shown in Figure \ref{f:deviation}(a). It is clear that even for $\bar{x}_{12}>>\bar{x}_{11}$ the standard deviation is small. Actually, the maximum value is observed for $\bar{x}_{12}\approx 3$, which indicates that for larger unbalance between communities the model should still provide good accuracy. The reason for the presence of a maximum deviation is not clear, and is a subject for future analyses in this model. 

\begin{figure}[!t]
    \includegraphics[width=0.95\linewidth]{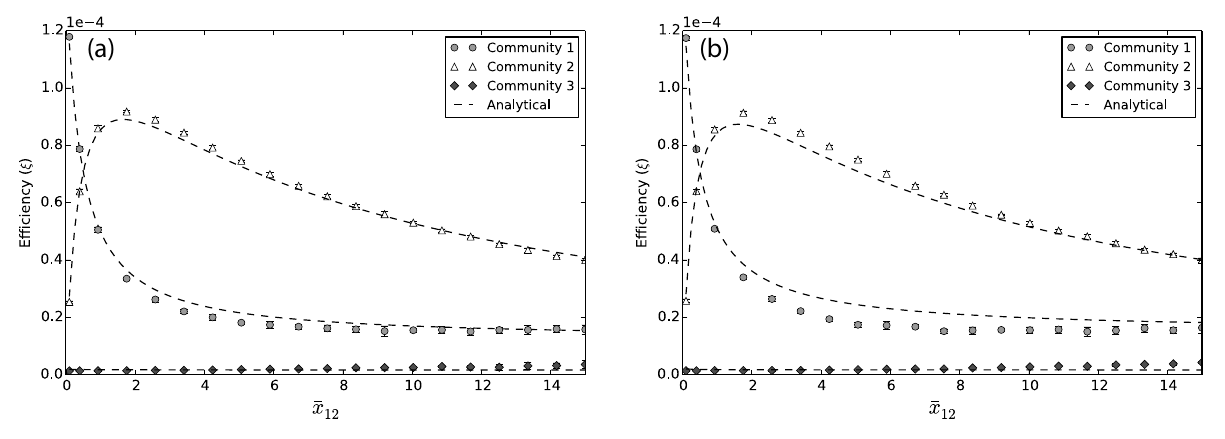}
    \caption{Efficiency variation when changing the balance between BA communities. (a) Comparison between experimental data, indicated by symbols, with the analytical prediction of Equation \ref{eq:prob_BA}, indicated by a dashed line. (b) Same as in (a), but using Equation \ref{eq:prob_BA_simple} for the analytical prediction.}
    \label{f:3_com_BA}
\end{figure}

Although Equation \ref{eq:prob_BA} can provide a good prediction of community efficiencies, it is overly complicated and might hinder important insights about the concept of efficiency. We can simplify the expression by considering that since the external degrees of the nodes are randomly assigned, they are well represented by the their average values, given by matrix $\kappa$. Therefore, we can rewrite Equation \ref{eq:prob_BA} as

\begin{equation}
    p_{mh}=a\underset{x_{mm}}{\sum}\frac{\bar{x}_{mh}(1-\delta_{m,h})+x_{mm}\delta_{m,h}}{x_{mm}^{3}\left(\underset{l\neq m}{\sum}\bar{x}_{ml}+x_{mm}\right)}.\label{eq:prob_BA_simple}
\end{equation}
In Figure \ref{f:3_com_BA} we show the mean efficiency of BA networks generated from our model using the matrix $\kappa$ indicated in Equation \ref{eq:kappa}. Each data point represents 100 realizations of the model. In Figures \ref{f:3_com_BA}(a) and (b) we compare the experimental data with Equation \ref{eq:prob_BA} and \ref{eq:prob_BA_simple}, respectively. We see that Equation \ref{eq:prob_BA} has good agreement with the experimental values. Equation \ref{eq:prob_BA_simple} also provides a fairly good prediction of each community efficiency. The difference between the experiment and the analytical curve in Figure \ref{f:3_com_BA}(a) around $\bar{x}_{12}\approx 3$ is caused by the larger standard deviation of the transition probabilities at this region (as shown in Figure \ref{f:deviation}(c)).

We note that Equations \ref{eq:prob_BA} and \ref{eq:prob_BA_simple} also work well when the sizes of the communities are distinct. Nevertheless, if the difference in sizes is too large (e.g., one community having hundreds of nodes and the other tens of thousands), one community will dominate the activity of the network. The analytical expressions will still provide a good estimation of the efficiency of the dominant community, but the smaller communities will present large fluctuations of activity, which are not well predicted by the model.

\subsection{Efficiency for the Macaque cortical network}

\begin{figure}[!t]
    \includegraphics[width=0.5\linewidth]{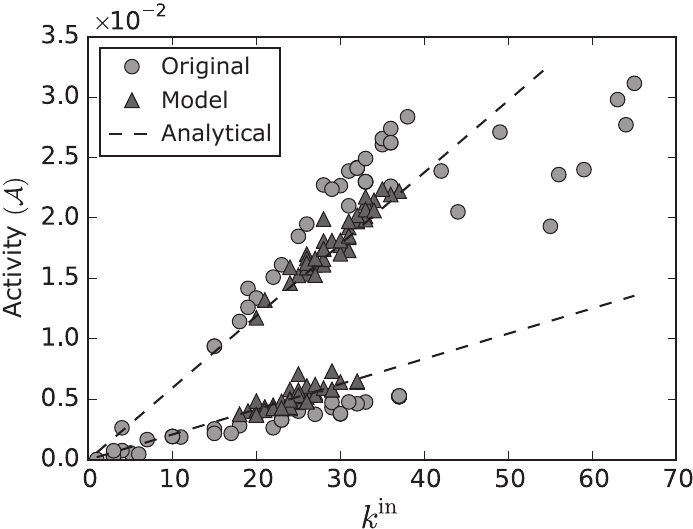}
    \caption{Activity of the cortical network (circles) compared to the model (triangles) and the analytical prediction (dashed lines). The average degree of the communities are $\bar{x}_{11} = 25.95$, $\bar{x}_{22}=23.91$, $\bar{x}_{12}=1.21$ and $\bar{x}_{21}=4.35$. The communities have 42 and 43 nodes.}
    \label{f:exp_cortical}
\end{figure}

We have obtained a reasonable model to explain the non-linearity observed in the relationship between activity and inward degree in the Macaque cortical network. Therefore, we found all the relevant parameters of each community previously detected in the cortical network and used them to perform a single realization of our model using a Poissonian distribution for the internal degrees of the communities. The result is shown in Figure \ref{f:exp_cortical}, where we can see that the model originates a two-group activity that is very similar to that found for the cortical network. In addition, since the network has only two communities, we can write Equation \ref{eq:activity_n} in a closed form for community 1 as

\begin{equation}
    {\cal A}_{1}=\frac{\pi_{1}}{\frac{N}{2}\left(\bar{x}_{21}+\bar{x}_{11}\right)}k_{i}^{\textnormal{in}},\label{eq:activity_2}
\end{equation}
where

\begin{equation}
\pi_{1}=\frac{1}{1+\frac{\bar{x}_{12}}{\bar{x}_{21}}\frac{\bar{x}_{21}+\bar{x}_{22}}{\bar{x}_{12}+\bar{x}_{11}}},\label{eq:P_1}
\end{equation}
with a similar expression holding for the second community. The predicted values of activity using Equation \ref{eq:activity_2} are shown in Figure \ref{f:exp_cortical}, where we see a good agreement between the analytical and original activity of the given cortical network. 


\begin{figure}[!t]
    \includegraphics[width=0.5\linewidth]{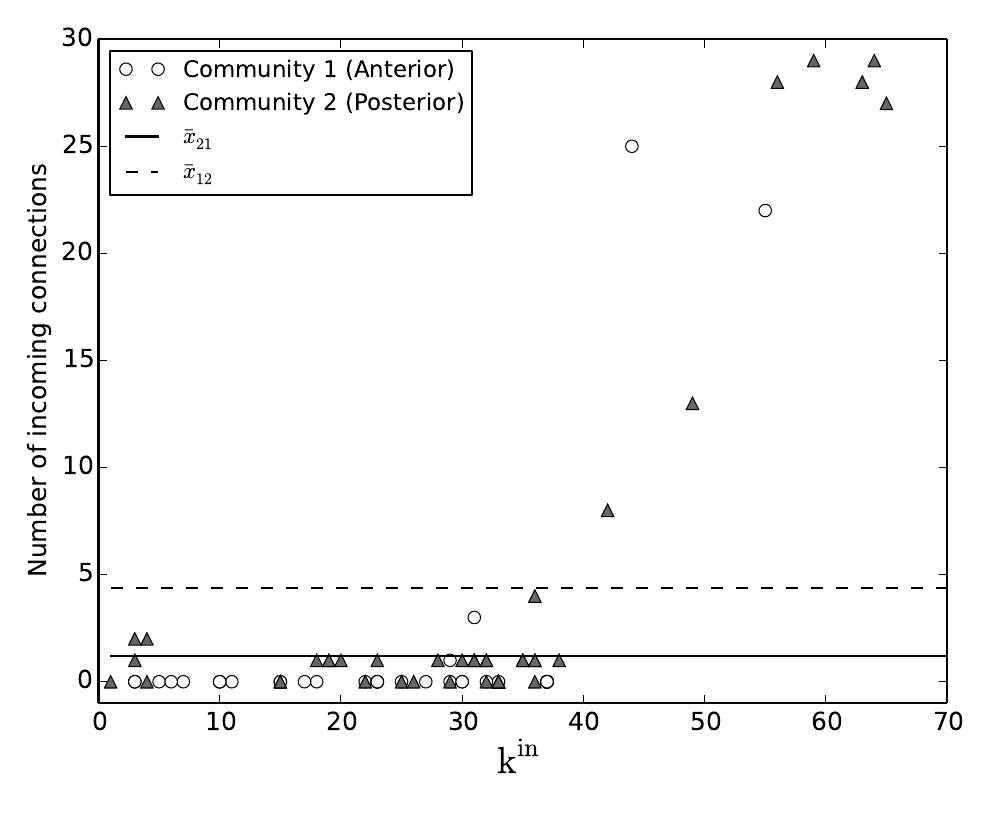}
    \caption{We plot, for each node, the number of inward connections belonging to other communities against the inward degree of the node. We also indicate with black lines the average number of incoming connections in the anterior (continuous line) and posterior (dashed line) communities.}
    \label{f:exp_deviation}
\end{figure}

A clear trend observed in Figure \ref{f:exp_cortical} is that nodes having large inward degree do not adhere to the model or the analytical curves. This is explained by a peculiar property displayed by such nodes. In Figure \ref{f:exp_deviation} we show the number of connections a node receives from the other community (posterior or anterior) as a function of its inward degree. It is clear that the large degree nodes tend to receive much more connections than other nodes in the same community. Such nodes serve as entry points to their respective communities, and therefore are influenced by the activity of both the anterior and posterior modules. Therefore, they tend to display an intermediate level of efficiency between the two communities. Also, in Figure \ref{f:exp_deviation} we indicate with solid and dashed lines the average values of external degrees used to generate the model network. The large degree nodes strongly influenced the obtained averages. As a consequence, they are responsible for the slight deviation of efficiencies obtained for the model and the analytical expression.

\section{Conclusions}

The loss of correlation in directed networks has been an intriguing phenomenon. We have shown in this article that this is intrinsically associated to the unbalance of directed connections between communities in a network. We also found that the communities topology and edges reciprocity do not significantly affect their respective efficiency. The derived equations provide the basis not only for better understanding the effect of modular structure in diffusion dynamics, but also for planning strategies for increasing or decreasing the number of accesses to nodes in real-world networks. For example, in the context of the WWW, if we want to increase the activity inside a given module, it may be more efficient to add connections coming from other modules than to increase the density of connections inside the module. The implications are particularly relevant for neuroscience, where the systems are almost invariantly modular and directed. It would also be interesting to study the efficiency along topological scales, as well as to extend the results reported in this work to other types of dynamics such as integrate and fire and synchronization.

\section*{Acknowledgments}
Luciano da F. Costa is grateful to FAPESP (12/50986-7) and CNPq (307333/2013-2) for the financial support. C.H.C. is grateful to FAPESP (11/22639-8) for sponsorship. M.P. Viana thanks to FAPESP for financial support (2010/16310-0).

\section*{References}

\bibliography{strucdyn}
\bibliographystyle{unsrt}

\end{document}